\documentclass[]{article}
\usepackage{amsmath,amsfonts,amssymb}
\usepackage{amsthm}
\usepackage{mathtools}
\usepackage{cases}
\usepackage[mathfrak]{}
\usepackage[latin1]{inputenc}
\usepackage[T1]{fontenc}
\usepackage{verbatim}
\usepackage{graphicx}
\usepackage{float}
\usepackage{mathrsfs}
\usepackage [all]{xy}
\usepackage{color}
\usepackage{tikz}
\usepackage{comment}
\usepackage{hyperref}
\usepackage{hyphenat}
\usepackage{authblk}

\newtheorem{theorem}{Theorem}[section]
\newtheorem{lemma}[theorem]{Lemma}
\newtheorem{proposition}[theorem]{Proposition}%

\newtheorem{remark}{Remark}%
\newtheorem{protocol}{Protocol}
\newtheorem{game}{Game}

\newtheorem{definition}{Definition}

\title{Semigroup action based on skew polynomial evaluation with applications to Cryptography}
\author[1]{Daniel Camaz\'on-Portela \footnote{The author was partially supported by grant PID2022-138906NB-C21 funded by MICIU/AEI/ 10.13039/501100011033 and by ERDF A way of making Europe.}}
\author[2]{Juan Antonio L\'opez-Ramos \footnote{The second author was partially supported by grant PID2022-140934OB-I00 funded by by MICIU/AEI/ 10.13039/501100011033 and by ERDF A way of making Europe.}}
\affil[1, 2]{Department of Mathematics, University of Almer\'ia, Carretera Sacramento, SN, Almer\'ia, 04120, Spain}
\date{}                     
\begin{document}
  \maketitle



\begin{abstract}
Through this work we introduce an action of the skew polynomial ring $\mathbb{F}_{q}\left[X; \sigma, \delta\right]$ over $\mathbb{F}_{q}$ based on its polynomial valuation and the concept of left skew product of functions. This lead us to explore the construction of a certain subset $\mathcal{T}(X)\subset\mathbb{F}_{q}\left[X; \sigma, \delta\right]$ that allow us to control the non-commutativity of this ring, and exploit this fact in order to build a public key exchange protocol that is secure in Canetti and Krawczyk model.
\end{abstract}

\section{Introduction}\label{sec1}

The study of semigroup actions constitutes a vast and very active area of mathematical research not only from a theoretical approach (see e.g. \cite{Erdal18}, \cite{Dikranjan20} and \cite{Cordeiro20}), but also from the applied side as actions of semigroups appear quite often as mathematical models of progressive processes. We can find examples of this fact in different areas, from computer science where plan executions are defined in terms of semigroup actions (see e.g. \cite{Pardo24}) to physics where a dynamical system can be seen as a semigroup action (see e.g. \cite{Rodrigues16}). 

One of the distinguishing features of semigroup and group actions comes from the fact that the non existence of inverses in the semigroup turns their orbits into non return routes. This characteristic of the semigroup orbits makes them much harder to analyse than the group orbits and make semigroup actions be an excellent algebraic platform in order to build potential post-quantum cryptography (PQC) candidates. On this direction, Maze et al. in \cite{Maze07} extend a generalization of the original Diffie-Hellman key exchange to the setting of a semigroup action ($G$\hyp action) on a finite set. Some other works that make use of these ideas are those of L\'opez-Ramos et al. \cite{Lopez17} and Gnilke et al. \cite{Gnilke24}. Moreover, some key exchange proposals based on this algebraic construction can be found in \cite{Castryck18}, \cite{Smith18} and \cite{Alamati20}. However, some recent cryptanalysis of protocols based on conmmutative semigroup actions (see e.g. \cite{Lopez24}) suggest that non-commutative semigroups should be considered in order to achieve security.

Skew polynomials rings were introduced by Oyestein Ore in the 1933 paper \cite{Ore33}. A skew polynomial ring $R=K\left[X; \sigma, \delta\right]$ is a ring of polynomials with non-commutative multiplication. When operating in a skew polynomial ring $R$, there are several key differences
from a commutative polynomial ring (see \cite{Bau19}). For example, factorizations are not unique, and a
polynomial may have more roots than its degree. Moreover, due to the non-commutativity,
there is a difference between left and right divisibility, resulting that evaluation is not as
straight-forward as in the commutative case. These are desirable properties for an algebraic structure in order to be used within coding and cryptographic purposes, what makes skew polynomial rings useful for coding \cite{Boucher09}, secret sharing \cite{Zhang10}, and even cryptographic key exchange \cite{Boucher10}.

In this work, we define an action of the skew polynomial ring $\mathbb{F}_{q}\left[X; \sigma, \delta\right]$ over $\mathbb{F}_{q}$ based on its polynomial valuation, and we exploit its structure in order to define an extended Diffie-Hellman key exchange. 

\section{Preliminaries}\label{sec2}

Firstly, we will introduce some algebraic concepts that are needed for the rest of this paper.

Let $K$ be a skew field and $X$ be a (nonempty) set on which the multiplicative group $K^{*}:=K\setminus\left\{0\right\}$ acts (on
the left). The action of $a\in K^{*}$ on $x\in X$ is denoted by $\varphi(a,x)$. We denote the set of all functions $f:X\rightarrow K$ by $\mathcal{F}(X)$. By abuse of notation, a constant function in $\mathcal{F}(X)$, whose value is $a\in K$, is simply denoted by $a$. Given functions $f,g:X\rightarrow K$, we let $f+g$ denote the pointwise addition of the functions $f$ and $g$.

\begin{definition}\cite[Definition 2.1.]{Aryapoor24} \label{LSP}
The left skew product of functions $f,g\in\mathcal{F}(X)$ is a function $f\ast g:X\rightarrow K$ defined as follows
\begin{equation}
(f\ast g)(x)=\begin{cases} f(\varphi(g(x),x))g(x) & \text{if $g(x)\neq 0$,} \\
                            0 & \text{if $g(x)=0$}.
														\end{cases}
\end{equation}
\end{definition}
It is easy to see that $a\ast f(x)=af(x)$, for every $a\in K$, $x\in X$. We shall henceforth denote $a\ast f$ simply by
$af$. Note that $(a,f)\rightarrow af$ turns $\mathcal{F}(X)$ into a left $K-$vector space. In the following lemma some properties of the skew product are collected.

\begin{lemma}\cite[Lemma 2.1.]{Aryapoor24} \label{lemSkewProd}
Let $f, h, g: X\rightarrow K$ be arbitrary functions. Then:
\begin{enumerate}
\item The constant function $1$ is a unit for $\ast$, that is, $f=f\ast 1=1\ast f$ .
\item $(f+g)\ast h=f\ast h+g\ast h$, that is, the right distributive law (with respect to pointwise addition) holds
for the skew product.
\item $(f\ast g)\ast h=f\ast (g\ast h)$, that is, $\ast$ is associative.
\end{enumerate}
\end{lemma}

It follows from this lemma that the set $\mathcal{F}(X)$ equipped with pointwise addition and the skew product is a structure known as ``right near-ring'' (see \cite{Pilz83}). We note that the skew product may not be left distributive with respect to pointwise addition.

\section{Semigroup action based on evaluation of skew polynomials}

\begin{definition}
A semigroup $(G,\cdot)$ is a set $G$ with an inner operation $\cdot: G\times G\rightarrow G$ that satisfies the associativity property i.e $\forall a,b,c\in G$ we have that
\begin{equation*}
(a\cdot b)\cdot c=a \cdot (b\cdot c).
\end{equation*}
\end{definition}

It follows from Lemma \ref{lemSkewProd} that $(\mathcal{F}(X),\ast)$  has a semigroup structure.

Now we can define semigroup action over a set $S$.
\begin{definition}
Let $S$ be a set, and $G$ a semigroup. A map $\phi:G \times S\rightarrow S$ is said to define a $G-$action if $\phi (a,\phi(b,s))=\phi(a\cdot b,s)$ for all $a,b \in G, s \in S$.
\end{definition}

\begin{lemma}
Let $\psi:\mathcal{F}(X)\times K\rightarrow K$ be the function defined by
\begin{equation}
\psi(f(X),b)=\varphi(f(b),b).
\end{equation}
Then $\psi$ defines an action (on the left) of $(\mathcal{F}(X),\ast)$ over $K$.
\end{lemma}

\begin{proof}
It is clear that $\psi(1_{\mathcal{F}(X)},b)=b$, so in order to prove that $\psi$ defines an action (on the left) of $\mathcal{F}(X)$ over $K$, it just remains to verify that for any $f(X),g(X)\in\mathcal{F}(X)$, $\psi((f\ast g)(X),b)=\psi(f(X),\psi(g(X),b))$. However, by Definition \ref{LSP} we have that
\begin{align*}
\psi((f\ast g)(X),b) & =\varphi((f\ast g)(b),b)=\varphi(f(\varphi(g(b),b))g(b),b), \\
                     & =\varphi(f(\varphi(g(b),b)),\varphi(g(b),b))=\psi(f(X),\varphi(g(b),b)), \\
										 & =\psi(f(X),\psi(g(X),b)).
\end{align*}
\end{proof}

From now on, we will consider $X=K$. If we denote by $\mathcal{P}_{d}(X)$ the set formed by polynomial functions of degree $d$, let $\mathcal{P}(X)=\bigcup_{d=1}\mathcal{P}_{d}(X)\subset\mathcal{F}(X)$, $\sigma:K\rightarrow K$ an endomorphism and $\delta:K\rightarrow K$ a $\sigma-$derivation. By considering the action $\varphi_{\mathcal{P}(X)}:K^{*}\times K\rightarrow K$, where $\varphi_{\mathcal{P}(X)}(a,b)=\sigma(a)ba^{-1}+\delta(a)a^{-1}$, we define the right near-ring $(\mathcal{P}(X),+,\ast)$, that we will denote by $\mathcal{K}\left[X; \sigma; \delta\right]$. Moreover, we define the corresponding action $\psi_{\mathcal{K}\left[X; \sigma; \delta\right]}: \mathcal{K}\left[X; \sigma; \delta\right]\times K\rightarrow K$, where $\psi_{\mathcal{K}\left[X; \sigma; \delta\right]}(P(X),a)=\varphi_{\mathcal{P}(X)}(P(a),a)$. \\

\begin{definition}\label{defW(X)}
Let us denote by $\left\{a_{1}, \ldots, a_{r}\right\}$ the elements of the orbit $\mathcal{P}_{d}\cdot a$, and $P_{ij}(X)=P_{ji}(X)$, for $i,j=1,\ldots, r$ with $i\neq j$, the polynomial functions of the form:
\begin{equation*}
P_{ij}^{\alpha}(a_{k})=\begin{cases} \alpha_{ij} & \text{if}\enspace k\neq i,j, \\
                            \alpha_{ij}^{i} & \text{if}\enspace k=i, \\
														\alpha_{ij}^{j} & \text{if}\enspace k=j,
\end{cases}
\end{equation*}
where $\alpha_{ij}, \alpha_{ij}^{i}, \alpha_{ij}^{j}\in K$, that verify 
\begin{equation}
\prod_{k}\alpha_{ik}^{i}\prod_{k\neq i}\alpha_{lk}=u_{il},
\end{equation}
\begin{equation}
\prod_{k}\alpha_{lk}^{l}\prod_{k\neq l}\alpha_{ik}=v_{il},
\end{equation}
 for some fixed $u_{il}, v_{il}\in K$. Let $u=\left(u_{12},\ldots,u_{1r},\ldots,u_{i,i+1},\ldots,u_{ir},\ldots,u_{r-1,r}\right)$ and \\ $v=\left(v_{12},\ldots,v_{1r},\ldots,v_{i,i+1},\ldots,v_{ir},\ldots,v_{r-1,r}\right)$. Then we define $\mathcal{W}^{(u,v)}$ as the product $\mathcal{W}^{(u,v)}=\prod_{i,j}P_{ij}^{\alpha}(X)$.
\end{definition}

\begin{remark}
Note that if $P(X)\in\mathcal{W}^{(u,u)}$ then $P(a_{i})=P(a_{j})$ for any $i,j=1,\ldots,r$.
\end{remark}

\begin{proposition}\label{PropAlmostInv}
If $P(X), Q(X)\in\mathcal{W}^{(u,v)}$, that is, $P(X)=\prod_{i,j}P_{ij}^{\alpha^{'}}(X)$ and $Q(X)=\prod_{i,j}P_{ij}^{\alpha^{''}}(X)$. Then, we have that for any $i,l=1,\ldots,r$
\begin{equation*}
P(a_{i})Q(a_{l})-P(a_{l})Q(a_{i})=0.
\end{equation*}
Moreover, if $P(X), Q(X)\in\mathcal{W}^{(u,v)}$ and $R(X), S(X)\in\mathcal{W}^{(u^{1},v^{1})}$, then for any $i,l=1,\ldots,r$ the following equalities hold:
\begin{align*}
(P+R)(a_{i})(Q+S)(a_{l})-(P+R)(a_{l})(Q+S)(a_{i}) & =0, \\
(P\cdot R)(a_{i})(Q\cdot S)(a_{l})-(P\cdot R)(a_{l})(Q\cdot S)(a_{i}) & =0.
\end{align*}
\end{proposition}

\begin{proof}
Firstly, as $P(X), Q(X)\in\mathcal{W}^{(u,v)}$, then we have that \small
\begin{align*}
& P(a_{i})Q(a_{l})-P(a_{l})Q(a_{i})=\prod_{j}\alpha^{i'}_{ij}\prod_{\substack{j\neq i \\ k}}\alpha^{'}_{jk}\prod_{j}\alpha^{l''}_{lj}\prod_{\substack{j\neq l \\ k}}\alpha^{''}_{jk}-\prod_{j}\alpha^{l'}_{lj}\prod_{\substack{j\neq l \\ k}}\alpha^{'}_{jk}\prod_{j}\alpha^{i''}_{ij}\prod_{\substack{j\neq i \\ k}}\alpha^{''}_{jk} \\
& =\left(\prod_{j}\alpha^{i'}_{ij}\prod_{k\neq i}\alpha^{'}_{lk}\prod_{j}\alpha^{l''}_{lj}\prod_{k\neq l}\alpha^{''}_{ik}-\prod_{j}\alpha^{l'}_{lj}\prod_{k\neq l}\alpha^{'}_{ik}\prod_{j}\alpha^{i''}_{ij}\prod_{k\neq i}\alpha^{''}_{lk}\right)\prod_{j, k\neq i,l}\alpha^{'}_{jk}\prod_{j, k\neq i,l}\alpha^{''}_{jk}, \\
& =\left(u_{il}v_{il}-v_{il}u_{il}\right)\prod_{j, k\neq i,l}\alpha^{'}_{jk}\prod_{j, k\neq i,l}\alpha^{''}_{jk}=0.
\end{align*} \normalsize
Now, as $R(X), S(X)\in\mathcal{W}^{(u^{1},v^{1})}$, then $R(X)=\prod_{i,j}P_{ij}^{\beta^{'}}(X)$ and $Q(X)=\prod_{i,j}P_{ij}^{\beta^{''}}(X)$. As a result, we have that \small
\begin{align*}
& (P+R)(a_{i})(Q+S)(a_{l})-(P+R)(a_{l})(Q+S)(a_{i})= \\
& \left(\left(\prod_{j}\alpha^{i'}_{ij}\prod_{j\neq i, k}\alpha^{'}_{jk}\right)+\left(\prod_{j}\beta^{i'}_{ij}\prod_{j\neq i, k}\beta^{'}_{jk}\right)\right)\left(\left(\prod_{j}\alpha^{l''}_{lj}\prod_{j\neq l, k}\alpha^{''}_{jk}\right)+\left(\prod_{j}\beta^{l''}_{lj}\prod_{j\neq l, k}\beta^{''}_{jk}\right)\right)- \\
& \left(\left(\prod_{j}\alpha^{l'}_{lj}\prod_{j\neq l, k}\alpha^{'}_{jk}\right)+\left(\prod_{j}\beta^{l'}_{lj}\prod_{j\neq l, k}\beta^{'}_{jk}\right)\right)\left(\left(\prod_{j}\alpha^{i''}_{ij}\prod_{j\neq i, k}\alpha^{''}_{jk}\right)+\left(\prod_{j}\beta^{i''}_{ij}\prod_{j\neq i, k}\beta^{''}_{jk}\right)\right)=  \\
& \left(u_{il}v_{il}-v_{il}u_{il}\right)\prod_{j, k\neq i,l}\alpha^{'}_{jk}\prod_{j, k\neq i,l}\alpha^{''}_{jk}+\left(u_{il}v^{1}_{il}-v_{il}u^{1}_{il}+u^{1}_{il}v_{il}-v_{il}^{1}u_{il}\right)\prod_{j, k\neq i,l}\alpha^{'}_{jk}\alpha^{''}_{jk}\prod_{j, k\neq i,l}\beta^{'}_{jk}\beta^{''}_{jk}+ \\
& \left(u^{1}_{il}v^{1}_{il}-v^{1}_{il}u^{1}_{il}\right)\prod_{j, k\neq i,l}\beta^{'}_{jk}\prod_{j, k\neq i,l}\beta^{''}_{jk}= 0;
\end{align*}
\begin{align*}
& (P\cdot R)(a_{i})(Q\cdot S)(a_{l})-(P\cdot R)(a_{l})(Q\cdot S)(a_{i})= \\
& \left(\prod_{j}\alpha^{i'}_{ij}\prod_{j\neq i, k}\alpha^{'}_{jk}\right)\left(\prod_{j}\beta^{i'}_{ij}\prod_{j\neq i, k}\beta^{'}_{jk}\right)\left(\prod_{j}\alpha^{l''}_{lj}\prod_{j\neq l, k}\alpha^{''}_{jk}\right)\left(\prod_{j}\beta^{l''}_{lj}\prod_{j\neq l, k}\beta^{''}_{jk}\right)- \\
& \left(\prod_{j}\alpha^{l'}_{lj}\prod_{j\neq l, k}\alpha^{'}_{jk}\right)\left(\prod_{j}\beta^{l'}_{lj}\prod_{j\neq l, k}\beta^{'}_{jk}\right)\left(\prod_{j}\alpha^{i''}_{ij}\prod_{j\neq i, k}\alpha^{''}_{jk}\right)\left(\prod_{j}\beta^{i''}_{ij}\prod_{j\neq i, k}\beta^{''}_{jk}\right)= \\
& \left(u_{il}u^{1}_{il}v_{il}v^{1}_{il}-v_{il}v^{1}_{il}u_{il}u^{1}_{il}\right)\prod_{j, k\neq i,l}\alpha^{'}_{jk}\prod_{j, k\neq i,l}\alpha^{''}_{jk}\prod_{j, k\neq i,l}\beta^{'}_{jk}\prod_{j, k\neq i,l}\beta^{''}_{jk}=0.														
\end{align*}
\end{proof}
\normalsize

\begin{definition}\label{defT(X)}
We define $\mathcal{T}(X)\subset\mathcal{K}\left[X; \sigma; \delta\right]$ as the subset $\mathcal{T}(X)=\mathcal{W}^{u^{0},v^{0}}\left(\oplus_{i=1}^{m}\mathcal{W}^{u^{i},v^{i}}\right)$, where $u^{0}=v^{0}$, that is, if $P(X)\in\mathcal{T}(X)$ then it is of the form
\begin{equation*} 
P(X)=P_{0}(X)\left(P_{1}(X)+P_{2}(X)+\ldots+P_{m}(X)\right),
\end{equation*}
with $P_{i}(X)\in\mathcal{W}^{u^{i},v^{i}}$ for $i=0,\ldots,m$.
\end{definition}
It follows from Proposition \ref{PropAlmostInv} that if $P(X), Q(X)\in\mathcal{T}(X)$ then $P(a_{i})Q(a_{l})=P(a_{l})Q(a_{i})$ for any $i,l=1,\ldots,r$.

\section{A key exchange protocol}\label{sec3}


Let us recall all the mathematical concepts that allow to get the formalization of the security model of \cite{Boh07}, developed from \cite{Kat03}. 

The (potential) participants in the protocol are modeled as probabilistic polynomial time (PPT) Turing machines in the finite set $\mathcal{U}=\{ \mathcal{U}_1,\dots ,\mathcal{U}_n\}$, and  every participant $\mathcal{U}_i$, $i=1, \dots ,n$ in the set $\mathcal{U}$ can run a polynomial amount of protocol instances in parallel.

We will denote by $\Pi_{U_{i}}^{i}$ to instance $i$ of participant $\mathcal{U}_i$ and we assign it the following variables:
\begin{itemize}
\item $pid_{U_{i}}^{i}$: stores the set of identities of those principals that $\Pi_{U_{i}}^{i}$ aims at establishing a key with, including $U_{i}$ himself.
\item $used_{U_{i}}^{i}$ : indicates whether this instance is or has been used for a protocol run.
\item $sk_{U_{i}}^{i}$: stores the session key once it is accepted by the instance $\Pi_{U_{i}}^{i}$. Before acceptance, it stores a distinguished $NULL$ value.
\item $sid_{U_{i}}^{i}$ : denotes a (non-secret) session identifier that can serve as identifier for the session key $sk_{U_{i}}^{i}$.
\item $acc_{U_{i}}^{i}$: indicates if the protocol instance was successful, that is, the session key is accepted.
\item $term_{U_{i}}^{i}$: shows if the execution has terminated.
\end{itemize}

\subsection{Cryptographic construction}

We are now in a position to introduce an extended Diffie-Hellman key exchange defined on a semigroup setting.
		
\begin{protocol}\label{protsemigroup}
Let $S$ be a finite field with $q=p^{m}$ elements, that is $S=\mathbb{F}_{q}$, $G$ the right near-ring $\mathbb{F}_{q}\left[X ; \sigma; \delta\right]$, $\psi$ the $G-$action over $S$ defined by $\psi(P(X),a)=\sigma(P(a))a(P(a))^{-1}+\delta(P(a))(P(a))^{-1}$ and $\mathcal{T}(X)\subset\mathbb{F}_{q}\left[X; \sigma; \delta\right]$ the subset of Definition \ref{defT(X)}.
    \begin{enumerate} 
        \item The initiator $\mathcal{U}_{i}$, on input $(\mathcal{U}_{i}, \mathcal{U}_{j}, sid_{U_{i}}^{i}=s)$, chooses a secret element $P(X)\xleftarrow{R}\mathcal{T}(X)$, and    sends $(\mathcal{U}_{i}, s, pk_{i}=\psi(P(X),a))$ to $U_{j}$.
        \item Upon receipt of $(\mathcal{U}_{i}, sid_{U_{j}}^{i}=s, pk_{i})$ the responder, $\mathcal{U}_{j}$ , chooses a secret element $Q(X)\xleftarrow{R}\mathcal{T}(X)$, and sends $(\mathcal{U}_{j}, s, pk_{j}=\psi(Q(X),a))$ to $\mathcal{U}_{i}$, computes $k_{j}=\psi(\widetilde{Q}(X),pk_{i})$, where $\widetilde{Q}(X)=\left(Q(pk_{i})+Q(pk_{j})\right)+\left(Q(pk_{i})+Q(pk_{j})\right)pk_{j}X+\ldots+\left(Q(pk_{i})+Q(pk_{j})\right)(pk_{j})^{d}X^{d}$, erases $Q(X)$ and outputs the key $sk_{U_{j}}^{i}=k_{j}$ under the session-id $sid_{U_{j}}^{i}=s$. 
        \item Upon receipt of $(\mathcal{U}_{j}, sid_{U_{i}}^{i}=s, pk_{j})$ the party, $\mathcal{U}_{i}$, computes $k_{i}=\psi(\widetilde{P}(X),pk_{j})$, where $\widetilde{P}(X)=\left(P(pk_{j})+P(pk_{i})\right)+\left(P(pk_{j})+P(pk_{i})\right)pk_{i}X+\ldots+\left(P(pk_{j})+P(pk_{i})\right)(pk_{i})^{d}X^{d}$, erases $P(X)$ and outputs the key $sk_{U_{i}}^{i}=k_{i}$ under the session-id $sid_{U_{i}}^{i}=s$.
    \end{enumerate}		
\end{protocol}

\begin{proposition} \label{secretKey}
Protocol \ref{protsemigroup} verifies correctness condition.
\end{proposition}
\begin{proof}
If both $\mathcal{U}_{i}$ and $\mathcal{U}_{j}$ are uncorrupted during the exchange of the key and both complete the protocol for session-id $s$, then they both establish the same key. Indeed, since the choice of $\widetilde{P}(X)$ and $\widetilde{Q}(X)$, we have $\widetilde{P}(X)\ast Q(X)(a)=\widetilde{Q}(X)\ast P(X)(a)$, then
\begin{align*}
k_{i} & =\psi(\widetilde{P}(X),pk_{j}) \\
      & =\varphi_{\mathcal{P}(X)}(\widetilde{P}\ast Q(a),a) \\
			& =\varphi_{\mathcal{P}(X)}(\widetilde{Q}\ast P(a),a) \\
			& =\psi(\widetilde{Q}(X),pk_{i})=k_{j}.
\end{align*}
\end{proof}

\subsection{Intractability assumptions}

We now present attack games for algebraic problems related to the security of our key exchange protocol.

\begin{game}[Semigroup Action Problem]
For a given adversary $\mathcal{A}$, we define
the following attack game:
\begin{itemize}
    \item The challenger computes:
    \begin{enumerate}
        \item $P(X)\xleftarrow{R}\mathcal{T}(X)$;
        \item $pk\leftarrow\psi(P(X),a)=\varphi_{\mathcal{P}(X)}(P(a),a)$;
    \end{enumerate}
    and gives the value $pk$ to the adversary $\mathcal{A}$.
    \item $\mathcal{A}$ outputs $\overline{P}(X)$.
\end{itemize}
\end{game}
We define $\mathcal{A}'$s advantage in solving the Semigroup action problem for $\mathbb{F}_{q}\left[X; \sigma; \delta\right]$,  denoted $SAPadv\left[\mathcal{A},\mathbb{F}_{q}\left[X; \sigma; \delta\right]\right]$, as the probability that $\varphi_{\mathcal{P}(X)}(P(a),a)=\varphi_{\mathcal{P}(X)}(\overline{P}(a),a)$.

\begin{definition}
We say that the Semigroup Action Problem (SAP) assumption holds for $\mathbb{F}_{q}\left[X; \sigma; \delta\right]$ if for all
efficient adversaries $\mathcal{A}$ the quantity $SAPadv\left[\mathcal{A},\mathbb{F}_{q}\left[X; \sigma; \delta\right]\right]$ is negligible.
\end{definition}

We now present other algebraic problems related to the SAP problem.

\begin{game}[Computational Generalized Semigroup Action Problem]
For a given adversary $\mathcal{A}$, we define
the following attack game:
\begin{itemize}
    \item The challenger computes:
    \begin{enumerate}
        \item $P(X)\xleftarrow{R}\mathcal{T}(X)$;
				\item $Q(X)\xleftarrow{R}\leftarrow\mathcal{T}(X)$;
        \item $pk_{1}\leftarrow\psi(P(X),a)=\varphi_{\mathcal{P}(X)}(P(a),a)$;
				\item $pk_{2}\leftarrow\psi(Q(X),a)=\varphi_{\mathcal{P}(X)}(Q(a),a)$;
				\item $k\leftarrow\psi(\widetilde{Q}(X),pk_{1})=\psi(\widetilde{P}(X),pk_{2})$;
    \end{enumerate}
    and gives the values of $pk_{1}$ and $pk_{2}$ to the adversary $\mathcal{A}$.
    \item $\mathcal{A}$ outputs $\overline{k}\in\mathbb{F}_{q}$.
\end{itemize}
\end{game}
We define $\mathcal{A}'$s advantage in solving the Computational semigroup action problem for $\mathbb{F}_{q}\left[X; \sigma; \delta\right]$, denoted $CGSAPadv\left[\mathcal{A},\mathbb{F}_{q}\left[X; \sigma; \delta\right]\right]$, as the probability that $\overline{k}=k$.

\begin{definition}[Computational Generalized Semigroup Action Problem Assumption]
We say that the Computational Generalized Semigroup Action Problem (CGSAP) assumption holds for $\mathbb{F}_{q}\left[X; \sigma; \delta\right]$ if for all efficient adversaries $\mathcal{A}$ the quantity $CGSAPadv\left[\mathcal{A},\mathbb{F}_{q}\left[X; \sigma; \delta\right]\right]$ is negligible.
\end{definition}

We now present a stronger computational problem,the decisional version of the $CGSAP$ problem.

\begin{game}[Decisional Generalized Semigroup Action Problem]
    For an adversary $\mathcal{A}$, we define the following two experiments: 
    \paragraph{Experiment $b$}\label{gameDCE}
    \begin{itemize}
    \item The challenger computes:
    \begin{enumerate}
        \item $P(X)\xleftarrow{R}\mathcal{T}(X)$;
        \item $Q(X)\xleftarrow{R}\mathcal{T}(X)$;
        \item $S(X)\xleftarrow{R}\mathcal{T}(X)$;
        \item $pk_{1}\leftarrow\psi(P(X),a)$; $pk_{2}\leftarrow\psi(Q(X),a)$;
        \item $k_{0}\leftarrow\psi(\widetilde{Q}(X),pk_{1})=\psi(\widetilde{P}(X),pk_{2})$; $k_{1}\leftarrow\psi(S(X),a)$;
    \end{enumerate}
    and gives the triple $(pk_{1}, pk_{2}, k_{b})$ to the adversary $\mathcal{A}$.
    \item $\mathcal{A}$ outputs a bit $\overline{b}\in\left\{0, 1\right\}$.
\end{itemize}
\end{game}
If $W_{b}$ is the event that $\mathcal{A}$ outputs $1$ in experiment $b$, we define $\mathcal{A}'$s advantage in
solving the Decisional Generalized Semigroup Action Problem for $\mathbb{F}_{q}\left[X; \sigma; \delta\right]$ as 
\begin{equation*}
DGSAadv=\left[\mathcal{A},\mathbb{F}_{q}\left[X; \sigma; \delta\right]\right]=\left|Pr\left[W_{0}\right]-Pr\left[W_{1}\right]\right|.
\end{equation*}

\begin{definition}[Decisional Generalized Semigroup Action Problem Assumption] We say that Decisional Generalized Semigroup Action Problem (DGSAP) assumption holds for $\mathbb{F}_{q}\left[X; \sigma; \delta\right]$ if for all efficient
adversaries $\mathcal{A}$ the quantity $DGSAPadv=\left[\mathcal{A},\mathbb{F}_{q}\left[X; \sigma; \delta\right]\right]$ is negligible.
\end{definition}

\subsection{Security analysis in the authenticated-links adversarial model}

Our first aim this section we aim to give a security proof of the key exchange protocol using the ideas of \cite{Canetti2001}. Thus, we need to fix the security model and this will be modeled using different oracles that are introduced in \cite{Kat03}. 

We are assuming that the adversary $\mathcal{A}$ has complete control over the network communications, but $\mathcal{A}$ cannot inject or modify messages,except for messages sent by corrupted parties or sessions. The different interactions that $\mathcal{A}$ can develop in the network can be described by the following oracles.
\begin{itemize}
\item Send($\mathcal{U}_{i}, M$): This sends the message $M$ to instance $\Pi_{U_{i}}^{i}$ of user $\mathcal{U}_i$. The output will be the protocol message that the instance outputs after receiving message $M$. The adversary $\mathcal{A}$ can use this protocol to initialize a protocol execution, by using the special message $M=\left\{\mathcal{U},i,\left\langle\mathcal{U}_{2},\ldots,\mathcal{U}_{n}\right\rangle\right\}$ to an unused instance.This oracle initializes a protocol run among users $\{\mathcal{U}_{2},\ldots,\mathcal{U}_{n}\} \subseteq \mathcal{U}$.

\item Execute($\mathcal{U}_1,\ldots,\mathcal{U}_n$): This executes the protocol between unused instances and returns a transcript of a complete execution of the protocol among the specified instances.

\item Reveal($\mathcal{U}, i$ ): this oracle returns the session key stored in $sk_{U_{i}}^{i}$.

\item Corrupt($\mathcal{U}_i$) : this query returns $\mathcal{U}_i$'s long term secret key.
\end{itemize}

According to \cite{Canetti2001}, in what is called the Authenticated-links Model (AM), an adversary $\mathcal{A}$ has absolute control on the communications links, by listening all the transmitted information, deciding what messages will reach their destination and when, although $\mathcal{A}$ cannot inject or modify any message. The adversary is also granted to obtain secret information stored in the parties memory that in by means of different attacks enumerated in \cite{Canetti2001}. All these capabilities could be seen as having access to the previously defined oracles, which is the setting we are going to consider.   

The aim of this section is to analyze the security of the Protocol \ref{protsemigroup} against an adversary with the capabilities defined in the Authenticated-links model introduced in \cite{Canetti2001}. In this case, the adversary is granted to have a single access to another oracle:

\begin{itemize}
\item Test$(\mathcal{U},i)$: after this query the oracle chooses $b\xleftarrow{R}\left\{0, 1\right\}$ and outputs the value of $sk_{U_{i}}^{i}$ in case $b=0$ and a random value of $\mathbb{F}_{q}$ if $b=1$. $\mathcal{A}$ can issue subsequent queries as desired, but the test session cannot be exposed.
\end{itemize}

Let $Guess\left\{\mathcal{A},\mathbb{F}_{q}\left[X; \sigma; \delta\right]\right\}$ be the event that $\mathcal{A}$ correctly guesses $b$, and define the advantage $SKadv\left\{\mathcal{A},\mathbb{F}_{q}\left[X; \sigma; \delta\right]\right\}=\left|Guess\left\{\mathcal{A},\mathbb{F}_{q}\left[X; \sigma; \delta\right]\right\}-\frac{1}{2}\right|$.

\begin{theorem}\label{thmsksecure}
If the $DGSA$ assumption holds for $\mathbb{F}_{q}\left[X; \sigma; \delta\right]$, then our key exchange
protocol is session-key secure in the authenticated-links adversarial model, that is $SKadv\left\{\mathcal{A},\mathbb{F}_{q}\left[X; \sigma; \delta\right]\right\}$ is negligible.
\end{theorem}

\begin{proof}
The proof is an adaptation to that one given by de la Cruz and Villanueva-Polanco \cite{delaCruz24}, developed from Canetti and Krawczyk original one \cite{Canetti2001}. We proceed by assuming to the contrary, i.e.there is an adversary $\mathcal{A}$ in the authentication-links model against our protocol that has a non-negligible advantage $\epsilon$ in guessing the bit $b$ chosen by the test oracle (when queried).Let $l$ be an upper bound on the number of sessions invoked by $\mathcal{A}$ in any interaction. We now construct a distinguisher $D$ for the $DGSA$ problem as shown next.
    
\noindent Input: $(a,\mathbb{F}_{q}\left[X; \sigma; \delta\right],pk_1,pk_2,k)$:
\begin{enumerate}
    \item  $r\xleftarrow{R} \left\{1, \ldots, \mathit{l}\right\}$.
    
    \item Invoke $\mathcal{A}$, on a simulated interaction with parties $\{\mathcal{U}_1,\dots ,\mathcal{U}_n\}$ except for the $r^{th}$ session. 
   
   \item For the $r^{th}$ session, $\mathcal{A}$ is provided with the output of Execute$(\mathcal{U}_i,\mathcal{U}_j)$ (this includes $pk_1$ and $pk_2$ sent by $\mathcal{U}_i$ and $\mathcal{U}_j$ respectively).
   
   \item If $\mathcal{A}$ chooses the $r^{th}$ session as the test session, then $\mathcal{A}$ is provided with the output of the query Reveal$(\mathcal{U},i)$:  $d \leftarrow \mathcal{A}(k)$.
   
   \item else $d \overset{R}{\leftarrow} \{0, 1\}$.

   \item return $d$. 
  
\end{enumerate}
Suppose that $\mathcal{A}$ cooses $r^{th}$ as the test session . Then $\mathcal{A}$ is provided with $k_{0}$, the session key stored in $sk_{U_{i}}^{i}=sk_{U_{j}}^{i}$ for $sid_{U_{i}}^{i}=sid_{U_{j}}^{i}=r$, or $k_{1}$ a random value in $(G, S, \phi)$. But $\mathcal{A}$ can distinguish it with a probability of $1/2 + \epsilon$, with $\epsilon$ non-negligible. Let us suppose now that $\mathcal{A}$ dos not choose the $r^{th}$ as the test session. Then the distinguisher $D$ outputs a random bit and thus the probability to distinguish the input correctly is $\frac{1}{2}$. Therefore, as the first case occurs with a probability of $\frac{1}{\mathit{l}}$, and the second case has a probability of $1-\frac{1}{\mathit{l}}$, then the probability for the distinguisher $D$ to win the $DGSA$ game is $\frac{1}{2}+\frac{\epsilon}{\mathit{l}}$, which is non-negligible.
\end{proof}

\begin{remark}
As proposed originally in \cite{Maze07}, the ``right near-ring'' structure of $\left(\mathcal{F}(X), +, \ast\right)$ allows us to define a probabilistic encryption scheme, which security is directly related to that of the key exchange protocol, that is:
\begin{enumerate}
\item $U_{i}$ chooses elements $P(X)\in\mathcal{T}(X)$ and $a\in\mathbb{F}_{q}$. $U_{i}'$ s public key is $(a, \psi(P(X),a))$.
\item $U_{j}$ chooses a random element $Q(X)\in\mathcal{T}(X)$ and encrypts a message $m$ using the encryption function 
\begin{equation*}
(m, Q(X))\rightarrow (\psi(Q(X),a), \psi(\widetilde{Q}(X),\psi(P(X),a))+m)=(c_{1}, c_{2}).
\end{equation*}
\item $U_{i}$ can decrypt the message using $m=-\psi(\widetilde{P}(X),c_{1})+c_{2}$.
\end{enumerate}
\end{remark}

\end{document}